\documentclass[useAMS,usenatbib,usegraphicx]{mn2e}
\usepackage{amsmath}
\usepackage{amssymb}
\usepackage{graphicx}
\usepackage{latexsym}
\usepackage{aas_macros}
\usepackage{natbib}
\usepackage[squaren]{SIunits}
\usepackage{array}
\usepackage{tabularx}

\newcommand{\scinot}[2]{\ensuremath{#1 \times 10^{#2}}}

\newcommand{\paren}[1]{\left ( #1 \right )}

\newcommand{\parenfrac}[2]{\paren{\frac{#1}{#2}}}

\newcommand{\differ}[1]{\textrm{d}#1}

\newcommand{\deriv}[2]{\frac{\differ{#1}}{\differ{#2}}}

\addunit{\cm}{\centi\metre}

\addunit{\erg}{erg}

\addunit{\persqcmnp}{\cm \rpsquared}
\addunit{\percubiccmnp}{\cm \rpcubed}
\addunit{\grampercubiccmnp}{\gram \usk \percubiccmnp}

\addunit{\yyear}{yr} 
\addunit{\AU}{au}

\newcommand{\mySun}{\odot}

\addunit{\Msol}{\ensuremath{\mathrm{M}_{\mySun}}}
\addunit{\Rsol}{\ensuremath{\mathrm{R}_{\mySun}}}
\addunit{\Lsol}{\ensuremath{\mathrm{L}_{\mySun}}}

\title[Radiative Feedback and Massive Star Formation]{A New Algorithm for Radiative Feedback and its Application to the  Formation of Massive Stars}
\author[R.G.~Edgar \& C.J.~Clarke]{Richard~Edgar$^1$\thanks{Email: rge21@ast.cam.ac.uk}
 and Cathie~Clarke$^1$ \\
$^1$Institute of Astronomy, Madingley Road, Cambridge CB3 0HA, Great Britain} 
\date{\today}

\pagerange{\pageref{firstpage}--\pageref{lastpage}} \pubyear{2002}

\begin{document}

\label{firstpage}

\maketitle

\begin{abstract}
  We have developed a simplified method of treating the radiative acceleration
  of dusty flows.
  This method retains the sharp impulse at the dust destruction radius that is
  a feature of frequency dependent radiative transfer, whilst placing minimal
  demands on computing resources.
  As such, it is suitable for inclusion in hydrodynamic codes.
  We have applied this method to the formation of massive stars in spherical
  geometry, and find that the fraction of a cloud which can accrete on to the
  central star is a strong function of the Jeans' Number and density profile of
  the cloud.
  Massive star formation is favoured by cold homogeneous conditions, as might
  result in regions where gas is swept up by some external triggering agent.
  We find (in contrast to previous authors) that massive star formation does not
  require a depleted dust mixture, although the use of dust at typical
  interstellar abundances does reduce the final stellar mass compared to cores
  formed from a depleted mixture.
\end{abstract}



\section{Introduction}

The question of massive star formation ($\gtrsim \unit{10}{\Msol}$) is
of great astrophysical interest.
Despite their rarity, massive stars dominate the light from young
star clusters.
As such, they are important tracers of star formation in distant
galaxies.
Within the Galaxy, massive stars also have profound effects on their
local environment.
During their brief lives, massive stars produce an intense UV
radiation flux, ionising and dispersing nearby gas
(see, e.g. \citet{1986MNRAS.221..635T}), while powerful
stellar winds are injected into the surroundings.
The death of a massive star is also a dramatic event - a supernova
which disperses metals into the interstellar medium, while also
injecting a large amount of mechanical energy.

Given this wide range of consequences, it is evidently desirable to
understand environmental effects that govern the formation of
massive stars.
Infall models for low mass stars have been studied for a number of
years.
Some of the earliest work in the area was performed (independently)
by \citet{1969MNRAS.145..271L} and \citet{1969MNRAS.144..425P}.
The possibility of magnetic fields was considered by many authors,
and ambipolar diffusion has been studied as a means of forming
the quasi-static systems often assumed by star formation simulations
(see, e.g. \citet{1979PASJ...31..697N} and
\citet*{1997ApJ...485..660S}).
Infrared excesses of slightly older protostars
(e.g. \citet{1968ApJ...151..977M}) are generally interpreted as
indicating the presence of accretion discs
(see \citet{1981ARA&A..19..137P}).
These ideas were drawn together by \citet*{1987ARA&A..25...23S},
and the infall model developed in this paper is now generally
accepted for low mass stars.
However, massive stars require a more sophisticated treatment.

The shorter evolutionary timescales of massive stars mean that
they are likely to join the main sequence before the end of their
main accretion phase.
This complicates simulations of the core itself - see
\citet*{1999MNRAS.310..360T} for how this can be a problem even for
low mass stars.
High accretion rates onto massive cores will also lead to large
luminosities, which may significantly affect the accretion flow
through radiation pressure on dust grains.
Moreover, since the main sequence lifetime of a massive star is
comparable to the free fall time of the parent core, it is evident
that the accretion history and luminosity evolution may be closely
linked throughout the star's life.
It should be noted that if stellar masses are self-limited by
radiation pressure on dust grains, then the high mass end of the IMF
is likely to be highly sensitive to the metallicity and dust grain
properties of the parent star forming region.
For more details, of these and other problems, see
\citet*{2000prpl.conf..327S}.

The possibility of radiation pressure halting accretion was first
considered by \citet{1971A&A....13..190L}.
Subsequent work by \citet{1986ApJ...310..207W,1987ApJ...319..850W}
concluded that stars with masses $> \unit{100}{\Msol}$ could form,
provided the dust abundance was depleted from normal galactic
values.
\citeauthor{1987ApJ...319..850W} considered spherically symmetric
accretion onto a massive star and modeled the interaction between
the stellar/accretion luminosity and the infalling dust.
Since interstellar dust typically melts when it reaches a temperature
in the range \unit{1500-2000}{\kelvin}, there is a significant amount
of momentum transfer in a region close to the dust destruction
radius, $r_{\textrm{d}}$, where the radiation from the central core first
impinges on the dusty inflow.
\citeauthor{1987ApJ...319..850W} used these models to propose
several criteria necessary for the formation of massive stars, of
which the two most important are
\begin{enumerate}
\item that accretion should proceed at a sufficiently high rate
that the ram pressure at $r_{\textrm{d}}$ exceeds the pressure associated with
the radiation impulse at that point
\item that the ratio of gravitational to radiative acceleration
at the outer edge should exceed unity, in order that the initially
static material will be set in motion in an inward direction
\end{enumerate}

These criteria have long been held to be inimical to the formation
of massive stars and have prompted a range of alternative models,
discussed below.
The first criterion requires that massive stars form in dense
cores
($n \sim \unit{10^6}{\percubiccmnp}$ for a \unit{100}{\Msol} core).
This was regarded as uncomfortable because if such case was in a
state of hydrostatic support prior to collapse (i.e. if the core
contained roughly a Jeans' Mass), the required temperature would
be high ($\sim \unit{1000}{\kelvin}$ for a \unit{100}{\Msol} core
according to \citeauthor{1987ApJ...319..850W},
which is much higher than the temperature in such regions -
according to section~3.3 of \citet{1999PASP..111.1049G}, temperatures
are typically $\sim\unit{100}{\kelvin}$, rising to a maximum of about
\unit{250}{\kelvin}).
The second criterion, on the other hand, translates (for a given
stellar mass and luminosity) into an upper limit on the permitted
value of the flux mean opacity at the outer edge of the core.
This turns out to be rather restrictive: for any reasonable value
of the radiation temperature at the outer edge, inflow is impossible
for the standard dust mixture of \citet*{1977ApJ...217..425M}.
Accretion is only permitted if the dust mixture is substantially
modified (reduction in the total mass of grains by a factor of four
from standard Galactic abundances - a factor of eight was used
in their simulations - and a smaller maximum size of the
graphite grains).
\citeauthor{1987ApJ...319..850W} thus concluded that `...special
grain conditions must exist for massive star formation to occur.'

These constraints have encouraged the discussion of alternative
models of massive star formation.
A straightforward and realistic modification of the above is to
allow for the angular momentum of the initial core and hence for
the breaking of spherical symmetry due to the development of a
centrifugally supported disc.
Yorke and collaborators (\citet{1979A&A....80..308Y,1980A&A....85..215Y},
\citet{1990ApJ...355..651B}, \citet*{1993ApJ...411..274Y},
\citet*{1995ApJ...443..199Y} and \citet{1999ApJ...525..330Y})
have pioneered the hydrodynamic study of such a situation, including
the effects of radiation pressure on dust.
They have pointed out that the problem of massive star formation
is considerably eased by the `flashlight' effect, whereby the bulk
of the ultraviolet photons escape along the rotation axis of the disc,
and thus do not intercept the accretion flow.
For reasons of computational economy, however, the majority of
such calculations have been performed using grey (frequency averaged)
opacities to compute the radiative acceleration of the flow.
\citeauthor{1987ApJ...319..850W} have shown that this simplification
may cause the efficacy of radiative feedback to be substantially
under-estimated (see below).
Very recently, \citet{2002ApJ...569..846Y} have produced the first
2.5D hydrodynamic simulations in which the acceleration due to
radiative feedback is computed using frequency dependent opacities.
The complexities of such a code, and the large computational overhead
it demands, has meant that it has not yet been possible to explore
a large area of parameter space.
Hence, it has not been possible to understand fully the
relationship between initial conditions and final stellar mass.
The continued increase in computational capacity should however bring
this problem within grasp in the next few years.

Alternatively, \citet*{1998MNRAS.298...93B} have suggested that
massive stars form through a mixture of accretion and coalescence
in the cores of ultra-dense clusters.
Coalescence is an attractive mechanism for massive star formation
inasmuch as the effect of radiative feedback is negligible.
Nevertheless, in such simulations massive stars also acquire mass
through Bondi--Hoyle type accretion
and it is unclear how radiative feedback operates for this accretion
geometry.

In this paper, we revisit the issue of spherical accretion onto
massive stars, through the use of hydrodynamical simulations
in which the algorithm for radiative feedback is designed to match
the outcome of frequency dependent radiative transfer calculations.
We compare the results of such simulations with those of
\citet{1987ApJ...319..850W}, who constructed \emph{steady} flows on
to stars of given mass and luminosity.
We find significant differences in the results, in the sense that
the conditions for massive star formation are apparently much less
restrictive than those envisaged by \citeauthor{1987ApJ...319..850W}:
in particular, we find there is no need to modify the grain mixture,
if collapse occurs from a highly Jeans unstable core.
Needless to say, spherical accretion is not a realistic model for
massive star formation, but it does represent the geometry in 
which radiative feedback is likely to be most effective (since both
accretion and radiative feedback occur over all solid angles).
Since we have shown that the formation of massive stars by accretion
is rather unproblematical even in spherical geometry, we may conclude
from this that the obstacles to massive star formation are still
less severe in the case of realistic (disc) geometries.

The structure of this paper is as follows:
Section~\ref{sec:SimpleRadTrans} discusses a simplified algorithm
for radiative feedback, which preserves the features of a full
frequency dependent treatment.
In section~\ref{sec:DynCode} we describe the hydrocode modified
to include the new algorithm.
Results are presented in section~\ref{sec:results}, and a detailed
discussion is in section~\ref{sec:discussion}.
Conclusions and closing remarks are in section~\ref{sec:conclude}.


\section{A New Algorithm for Radiative Feedback}
\label{sec:SimpleRadTrans}

One of the most important findings of the radiative transfer calculations of
\citet{1987ApJ...319..850W} was that the grey approximation is completely
inadequate to describe the physics involved in the dusty inflow.
The problem may be seen by comparing figures~6 and~7 of
\citet{1987ApJ...319..850W} - the calculations which used the Rosseland
mean opacity completely failed to produce the sharp deceleration at the
inner edge of the dust shell.
This point is also made by \citet*{1995A&A...299..144P}.

The problem stems from the extreme manner in which the dust opacity varies
with wavelength - typically $\kappa \propto \lambda^{-1.5}$.
Use of the Rosseland mean opacity carries an implicit assumption of
thermal equilibrium between the radiation and the fluid.
At the inner edge of the dust shell, the dust (melting at
$\sim\unit{2000}{\kelvin}$) encounters the radiation field coming
from the stellar surface (at a radiation temperature of
$\sim\unit{\scinot{2}{4}}{\kelvin}$).
The assumption of thermal equilibrium is evidently extremely poor
at this point in the flow, and the steep variation of the dust opacity
with wavelength ensures that this mismatch appears in the dynamics.
Figure~8 of \citet{1986ApJ...310..207W} implies that the Rosseland
optical depth of the envelope is no more than about 2, but it is
easy to show that the optical depth at ultraviolet wavelengths is on
the order of hundreds.
The use of Rosseland opacities therefore artificially softens the
impact of the radiation field.


\subsection{The Simplified Model}

In order to remedy this situation, we here develop a simplified algorithm
which retains the sharp impulse at the inner edge of the dust shell (a
feature of the full radiative transfer calculations), whilst at the same
time effecting considerable computational economies.
The prescription obtained is suitable for incorporation into hydrodynamics
codes.


\subsubsection{Radiative Transfer}

The essence of the method is to follow \citet{1986ApJ...310..207W} in
splitting the radiation field into two components: the direct
stellar radiation field ($L_*$) and a diffuse, thermalised field
($L_{\textrm{th}}$).
The total luminosity, comprised of the nuclear luminosity of the
core and any accretion luminosity (which both change as the protostar
evolves), is conserved, so that
\begin{equation*}
L_{\textrm{tot}} = L_* + L_{\textrm{th}}
\end{equation*}
at all times and at all radii (when integrated over frequency).
The thermalised radiation field may be treated using grey opacities,
since figures~6 and ~7 of \citet{1987ApJ...319..850W} show that deep
within the dust shell, this approximation is valid.

The thermalised radiation field was calculated as follows:
At every radius, a photospheric temperature could be defined via
\begin{equation}
L_{\textrm{th}}(r) = 4 \pi r^2 \sigma T_{\textrm{rad}}^4
\label{eq:SteffBoltzLaw}
\end{equation}
The radiation temperature thus obtained is that required to transport
the radiative flux, \emph{if the radiation became free streaming at
that point}.
The optical depth to a potential photosphere at radius $r$ may be
calculated as
\begin{equation}
\tau(r) = - \kappa_{\textrm{R}} \paren{T_{\textrm{rad}}(r)} \int_{\infty}^{r} \rho(r') \, \differ{r'}
\label{eq:RosselandPhotosphere}
\end{equation}
where, $\kappa_{\textrm{R}}$ is the Rosseland opacity, and for simplicity,
it is assumed that no significant reprocessing of radiation occurs
outside the potential photosphere.
The location of the actual photosphere for the dust shell may then
be found by the conventional definition: a photosphere lies at an
optical depth of $2/3$ from infinity.

Outside the photosphere, the value of $T_{\textrm{rad}}$ may be set equal
to the radiation temperature of the photosphere, as per the assumption
made by equation~\ref{eq:RosselandPhotosphere}.
Inward of the photosphere, the diffusion approximation must be
solved:
\begin{equation}
\deriv{T_{\textrm{rad}}}{r} =
- \frac{3 \kappa_{\textrm{R}} \rho}{4 \pi \sigma r^2 T_{\textrm{rad}}^3} L_{\textrm{th}}(r)
\label{eq:RadiativeDiffuseApprox}
\end{equation}
Since the grains are flowing inwards, it is reasonable to assume that
they will melt once the radiation temperature reaches
$T_{\textrm{sub}} \sim \unit{2000}{\kelvin}$, permitting the inner edge of the dust
shell to be located (see below also).

There remains the question of calculating $L_{\textrm{th}}(r)$.
This can be done by attenuating the stellar radiation
field as it proceeds outwards from $r_d$ via
\begin{equation}
L_{\lambda}^{*}(r) = L_{\lambda}^{*}(r_{\textrm{d}}) e^{-\tau_{\lambda}}
\label{eq:StellarLumAttenuate}
\end{equation}
where
\begin{equation}
\tau_{\lambda} = \int_{r_{\textrm{d}}}^{r} k^{\textrm{pr}}_{\lambda} \rho \, \differ{r}
\end{equation}
is the wavelength-dependent optical depth of the inflow, and
$k^{\textrm{pr}}_{\lambda}$ is the wavelength dependent radiation pressure
opacity, as defined by \citeauthor{1987ApJ...319..850W}.
There is the potential for this to have an unpleasant effect on
equation~\ref{eq:RadiativeDiffuseApprox}, by affecting $T_{\textrm{rad}}$.
In this case, a slow iterative solution would be required.
However, as outlined below, this problem was avoided in the simulations
presented.

The mechanical effect of the radiation field is provided
using
\begin{equation}
a_{\textrm{rad}}(r) = \frac{\int k^{\textrm{pr}}_{\lambda} L_{\lambda}(r) \, \differ{\lambda}}
                                    {4 \pi r^2 c}
\label{eq:FullRadiationImpulse}
\end{equation}
which may be calculated separately for the direct and thermalised fields.
Since the thermalised field is assumed to be a black body, the integral
over wavelength may be calculated in a lookup table of Planck mean
opacities.


\subsubsection{Dust Physics}

The dust physics were brutally pruned, compared to the model of
\citet{1987ApJ...319..850W}.
Instead of following the size evolution of the dust as it was destroyed, the dust
is taken to be present or absent, based on its equilibrium temperature
in the radiation field.
This may be calculated using
\begin{equation}
\frac{1}{(4 \pi r_{\textrm{d}})^2} \int_0^{\infty} Q^{\textrm{A}}(a,\lambda) L^*_{\lambda} \, \differ{\lambda}
=
\int_0^{\infty} Q^{\textrm{A}}(a,\lambda) B_{\lambda}(T_{\textrm{sub}}) \, \differ{\lambda}
\label{eq:DustTemperatureDetermine}
\end{equation}
where $Q^{\textrm{A}} (a, \lambda)$ is the absorption efficiency as a function
of wavelength for grains of radius $a$.
The desired sublimation temperature, $T_{\textrm{sub}}$, grain radius, and grain type
may be selected appropriately.

Melting the dust at a set temperature (and, implicitly, having it spontaneously
reform at that temperature if outflow occurs) is a gross simplification
to the true physical conditions.
However, this assumption is often used - see, e.g.
\citet{1990ApJ...355..651B,1993A&A...279..577P,1997ApJ...486..372B}.
It has been calculated in studies of low mass stars
(e.g. \citet*{1994A&A...290..573K}) that, once conditions are favourable for the
formation of dust, the grains grow rapidly.
The caveat is the assertion of \citet{1999A&A...347..594G} that nucleation is
a rather difficult process.
The appropriate destruction temperature is a matter of some debate.
The temperature of \unit{2000}{\kelvin} conventionally quoted is based on the
sublimation of graphite grains.
However, the dust destruction temperatures quoted by \citet*{1995ApJ...447..848L}
are much lower - typically in the region of \unit{1200}{\kelvin}.
\citeauthor{1995ApJ...447..848L} suggest that the discrepancy is caused by
mistaken assumptions about dust destruction processes.
However, \citeauthor{1995ApJ...447..848L} found that graphite grains tended to
be destroyed by chemispluttering (collisions with hydrogen leading to the formation
of short hydrocarbon chains).
Only silicates were destroyed by sublimation - and at a much lower temperature.
\citet*{1996A&A...312..624D} present similar calculations for a protoplanetary disk.
They also found that graphite grains were destroyed at relatively low temperatures
by surface reactions, while the silicates sublimated.
However, their silicate grains survive to $\sim \unit{2100}{\kelvin}$.
In the dynamical collapses presented, $T_{\textrm{sub}} = \unit{1700}{\kelvin}$
was used.

In the present work, all dust opacities were calculated from the efficiency
tables of Draine.
These tables are based on the work of \citet{1984ApJ...285...89D},
\citet{1993ApJ...414..632D} and \citet{1993ApJ...402..441L}, and are updated
versions of the values used by \citeauthor{1986ApJ...310..207W}.
The dust size distribution was taken to be the standard power law of
\citet{1977ApJ...217..425M}.


\subsubsection{Stellar Radiation}

In order to perform the radiative transfer calculation, it is necessary
to specify the radiation incident on the inner edge of the dust shell.
The spectrum of the stellar radiation field was modified in the manner of
\citet{1986ApJ...310..207W}, with radiation more energetic than the Lyman
Limit reprocessed to lie below it.
That is, the stellar spectrum, $B^{*}_{\lambda}$ is chosen such that
\begin{equation}
B^{*}_{\lambda} = 
\begin{cases}
0 & \text{for $0 < \lambda < 912 \, \mbox{\AA}$} \\
\epsilon B_{\lambda} & \text{for $\lambda > 912 \, \mbox{\AA}$}
\end{cases}
\label{eq:StellarSpectrum}
\end{equation}
and
\begin{equation}
\int_0^{\infty} B_{\lambda} \differ{\lambda} =
\int_0^{\infty} B^{*}_{\lambda} \, \differ{\lambda}
\label{eq:RadiateConserveEnergy}
\end{equation}
where $B_{\lambda}$ is a true black body spectrum, and $\epsilon$ is chosen to
ensure equation~\ref{eq:RadiateConserveEnergy} remains true.
Then,
\begin{equation*}
L^*_{\lambda} = 4 \pi^2 B^{*}_{\lambda} R_*^2
\end{equation*}


\subsection{Testing the Simplifications}

The approximations described above were tested against the full radiative
transfer calculations of \citet{1987ApJ...319..850W}.
The reference model was the \unit{100}{\Msol} calculation of
\citet{1987ApJ...319..850W}.
This model had a steady accretion rate of \unit{\scinot{5}{-3}}{\Msol\usk\reciprocal\yyear}
and a total core luminosity of \unit{\scinot{2.43}{6}}{\Lsol}.
Where possible, boundary conditions were taken to be those specified by
\citeauthor{1987ApJ...319..850W}, so equation~\ref{eq:DustTemperatureDetermine}
was not used.
Use of these boundary conditions also eliminated the troublesome relation between
equations~\ref{eq:RadiativeDiffuseApprox}
and~\ref{eq:StellarLumAttenuate}, since the dust would always be destroyed
in the innermost grid cell.

The gas equation of motion was modified to be
\begin{equation}
\frac{1}{2} \deriv{u^2}{r} = -\frac{G M(r)}{r^2} - \frac{1}{\rho}\deriv{P}{r}
                             + \frac{\int k^{\textrm{pr}}_{\lambda} L_{\lambda}(r) \, \differ{\lambda}}
                                    {4 \pi r^2 c}
\label{eq:ClothoGasEquation}
\end{equation}
where $M(r)$ is the mass enclosed at radius $r$, and $P$ is the gas pressure.
The $\lambda$ subscripts indicate quantities that are wavelength dependent.
This is based on equation~31 of \citet{1987ApJ...319..850W}.
However, the left hand side has been changed into a ram pressure gradient, while the
dust to gas mass ratio (which is $\sim 10^{-5}$) has been dropped from the right
hand side.

\begin{figure*}
\centering
\includegraphics[scale=0.8]{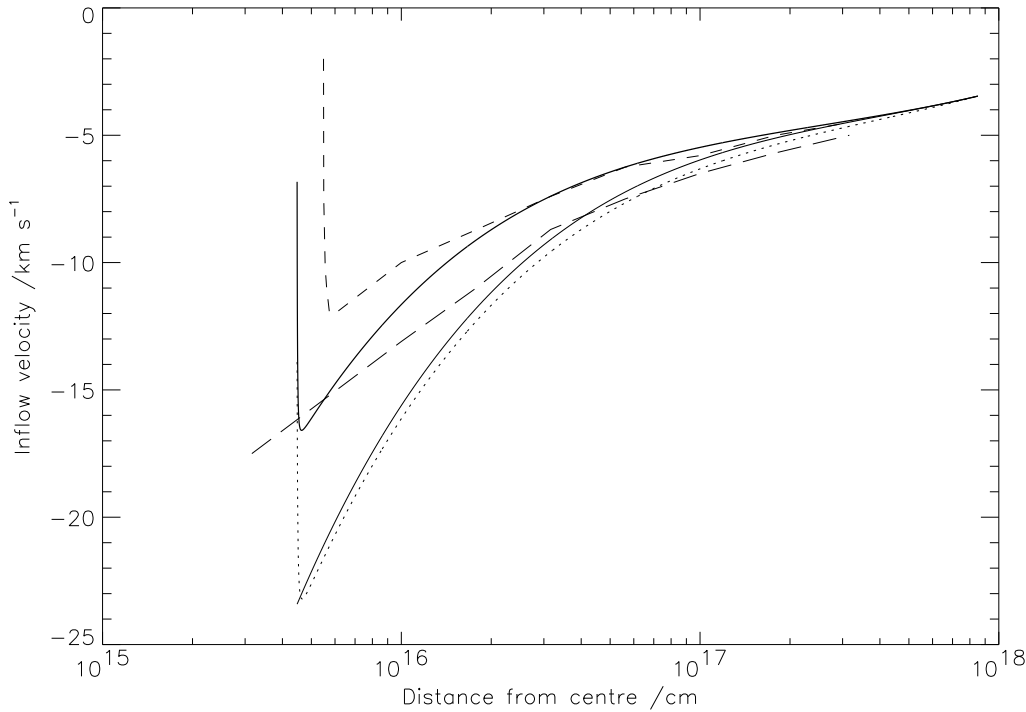}
\caption{Inflow velocities onto \unit{100}{\Msol} core. 
The thick solid curve is that obtained from the new algorithm,
while the thin solid curve is the freefall solution.
The dashed curves are points read from \citet{1987ApJ...319..850W}, with
the short dashes following the full solution of their figure~6, while the long
dashes follow the Rosseland mean calculation of their figure~7.
The dotted line is the curve one obtains if the effects of $L_{\textrm{th}}$
are omitted in the new algorithm}
\label{fig:100MsolSteadyInflowCurves}
\end{figure*}

The steady state velocity field obtained is plotted in
Figure~\ref{fig:100MsolSteadyInflowCurves}.
It may be seen that, although the general behaviour has been duplicated, there
is a discrepancy of a few kilometres per second in the inner portions of the
flow.
This difference can probably be ascribed to the extremely approximate nature
of the radiative transfer method used.
Since the dust shell has an optical depth of the order unity at temperatures
appropriate to the diffuse field, it is inevitable that errors will creep in - this
regime is notoriously tricky to approximate.
While it is certain that some method could be concocted to improve this specific
case, its generality would be suspect.
It should be noted that the magnitude of the jump in the inflow velocity at the
inner edge of the dust shell is identical to that of \citet{1987ApJ...319..850W}.
Comparing the momentum fluxes of the curves, the initial encounter with the stellar
radiation gives a change equivalent to $0.96 L/c$, while the thermalised field adds
another $0.71 L/c$.
These numbers are as one might expect - all the luminosity\footnote{
The missing four percent is probably due to the changing velocity affecting
the density structure (since the accretion rate is constant), which will then
change the gravitational forces slightly
}
is absorbed and re-emitted at the inner edge of the dust shell.
However, the thermal processing is significant, contributing almost as much
momentum flux as the initial strike.
Again, this is as one would expect, given that the entire shell has
$\tau_{\textrm{R}} \sim 1$.

The algorithm presented is a substantial improvement over calculations using
only the Rosseland mean (which cannot even reproduce the \emph{qualitative}
behaviour of the inflow).
Although there is a discrepancy of a few $\kilo\metre\usk\reciprocal\second$
at the inner edge of the dust shell, we note that the regime under
consideration was the most problematic to approximate, with Rosseland optical
depths only slightly larger than unity.
We would expect the approximations we have made to become better as the dust
shell becomes optically thicker.


\section{The Dynamics Code}
\label{sec:DynCode}

The {\sc ZEUS-2D} code of \citet{1992ApJS...80..753S} was chosen to simulate the
collapses.
Since the flows were generally found to be highly supersonic, and fairly
optically thin in the Rosseland mean, the gas equation of state
was taken to be isothermal, with the temperature set at the start of
the simulation.

Resolution considerations lead to a slight modification of the algorithm
detailed in section~\ref{sec:SimpleRadTrans}.
The loop between equations~\ref{eq:RadiativeDiffuseApprox}
and~\ref{eq:StellarLumAttenuate} was broken by noting that the flow is
likely to be so optically thick that the direct stellar field will be
attenuated in a single grid cell.
In this case, $L_{\textrm{th}}(r) = L_{\textrm{tot}}$ outside this grid cell.
The mechanical impulse exerted on the inflow at the dust destruction
radius may be represented by the injection of $L_{\textrm{tot}}/c$ of
momentum into the dust destruction grid cell.
If a photosphere is not found according to
equation~\ref{eq:RosselandPhotosphere}, then the value of $r_{\textrm{d}}$
may be calculated using equation~\ref{eq:DustTemperatureDetermine}
and the stellar field swept outwards from this point using
equation~\ref{eq:StellarLumAttenuate}.
No thermalised field need be calculated in this case, since its effect will
be minimal.

To avoid the central grid cell reaching unreasonable densities (and thereby
consuming impractical amounts of computer time), {\sc ZEUS} was modified so that
a linearly decreasing amount of mass was accreted from the innermost grid
cells, and placed into a point mass at the centre of the grid.
This simulated the formation of the protostar - which is not resolved in
our work.
The number of cells which were permitted to accrete, and the maximum fraction
accreted were found to have little effect on the results.
The point mass was used to determine the intrinsic luminosity and the radius
(for the accretion luminosity) of the forming protostar.
These quantities were determined using the Zero Age Main Sequence (ZAMS)
formulae of \citet{1996MNRAS.281..257T}.\footnote{Note that this assumption
will overestimate the luminosity of the core, and increase the effects of
radiative feedback}
To allow for the fact that the core would not immediately settle to its main
sequence radius, the method of \citet{1979A&A....80..308Y} was adopted:
once the core reached \unit{0.1}{\Msol}, its radius was set equal to
that of the first grid cell.
It was then allowed to contract on its instantaneous Kelvin--Helmholtz timescale,
until it shrank to its main sequence radius.
It is of course undesirable for the physical behaviour to be determined
by computational considerations (the positioning of the grid cell).
However, tests at different resolutions found no significant variation
in behaviour due to this.

Most runs were performed with a grid of 600 cells, fifty of which were
concentrated in the inner \unit{100}{\AU} with linear spacing.
The remainder were logarithmically spaced.
Higher resolution runs were also performed, typically with 800 grid cells
following the same general distribution pattern.


\section{Results}
\label{sec:results}

Two sets of simulations were performed.
The first started from uniform initial conditions, the second
started from a power law density profile.


\subsection{Collapse from Uniform Initial Conditions}

In Tables~\ref{tbl:Results241MsolCollapse}, \ref{tbl:Results88MsolCollapse}
and~\ref{tbl:Results5630MsolCollapse} we detail a number of simulations
which start from uniform initial density.
The initial density was \unit{10^{-19}}{\grampercubiccmnp}, with the
different masses obtained by varying the simulation volume.
The temperature of the gas, $T_{\textrm{gas}}$ of each run was fixed at the
indicated temperature; the corresponding value of $\alpha$ is also listed,
where $\alpha$ was defined as
\begin{equation}
\alpha = \frac{E_{\textrm{th}}}{|E_{\textrm{grav}}|}
\label{eq:DefineAlpha}
\end{equation}
where $E_{\textrm{grav}}$ is the system gravitational energy and
\begin{equation*}
E_{\textrm{th}} = \frac{3 \mathcal{R} M T_{\textrm{gas}}}{2 \mu}
\end{equation*}
where $\mathcal{R}$ is the molar gas constant.\footnote{Note that with
this definition, $\alpha$ is related to the Jeans' Number, $N_{\textrm{J}}$
by $N_{\textrm{J}} = \alpha^{-\frac{3}{2}}$}
Note that by setting $c_{\textrm{V}} = \frac{3}{2}\mathcal{R}$ the freeze-out of the
rotational modes of molecular hydrogen is ignored.
However, this is appropriate to the isothermal nature of the simulations.
Unless stated otherwise, the dust mixture is the depleted one employed by
\citet{1987ApJ...319..850W}.
The `normal' dust mixture is that of~\citet{1984ApJ...285...89D}, which is
used by \citet{1986ApJ...310..207W}.
Some of the simulations had no thermalised radiation field present
(that is, $L_{\textrm{th}} = 0$ everywhere), as indicated.
It should be noted that these runs lead to higher final stellar masses than
their counterparts with the full feedback mechanism, pointing to the
importance of the reprocessed radiation.
Variation of the grid resolution was not found to have a significant impact
on the final stellar masses.

\begin{table}
\centering
\begin{tabular}{c|c|c|>{\raggedright}p{3cm}}
$T_{\textrm{gas}} / \kelvin$  &  $\alpha$  &  $M_* / \Msol$   &  Comments \tabularnewline
\hline
10  &  0.03 & 196  & No thermalised field \tabularnewline
200 &  0.54 &  88  & No thermalised field \tabularnewline
 10  &  0.03 & 185.7  & \tabularnewline
 10  &  0.03 & 188.3  & High resolution \tabularnewline
 10  &  0.03 & 183.8  & High resolution \tabularnewline
 20  &  0.06 & 172.6  & \tabularnewline
 50  &  0.14 & 126.5  & \tabularnewline
100  &  0.27 & 84.4  & \tabularnewline
150  &  0.41 & 67.1   & \tabularnewline
200  &  0.54 & 48.7   & \tabularnewline
200  &  0.54 & 47.7   & High resolution \tabularnewline
300  &  0.81 & 39.9   & \tabularnewline
400  &  1.08 & 45.1   & \tabularnewline
 10  &  0.03 & 137.1  & Normal dust levels \tabularnewline
 10  &  0.03 & 138.4  & Normal dust levels, high resolution \tabularnewline
 20  &  0.06 & 106.4  & Normal dust levels \tabularnewline
 50  &  0.14 & 54.5   & Normal dust levels \tabularnewline
100  &  0.27 & 15.8   & Normal dust levels \tabularnewline
150  &  0.41 & 13.6   & Normal dust levels \tabularnewline
200  &  0.54 & 12.7   & Normal dust levels \tabularnewline
300  &  0.81 & 12.3   & Normal dust levels \tabularnewline
400  &  1.08 & 12.4   & Normal dust levels \tabularnewline
400  &  1.08 & 11.7   & Normal dust levels, high resolution \tabularnewline
\end{tabular}
\caption{Results from \unit{241}{\Msol} collapses of uniform gas}
\label{tbl:Results241MsolCollapse}
\end{table}

\begin{table}
\centering
\begin{tabular}{c|c|c|>{\raggedright}p{3cm}}
$T_{\textrm{gas}} / \kelvin$  & $\alpha$ & $M_* / \Msol$   &  Comments \tabularnewline
\hline
10   &  0.05 & 64.0  & \tabularnewline
20   &  0.11 & 56.0  & \tabularnewline
50   &  0.27 & 40.5  & \tabularnewline
100  &  0.54 & 30.2  & \tabularnewline
150  &  0.81 & 26.5  & \tabularnewline
200  &  1.08 & 26.8  & \tabularnewline
300  &  1.62 &  -    & Did not collapse \tabularnewline
 10   & 0.05 & 37.6 & Normal dust levels \tabularnewline
 20   & 0.11 & 23.7 & Normal dust levels \tabularnewline
 50   & 0.27 & 12.2 & Normal dust levels \tabularnewline
100   & 0.54 & 11.5 & Normal dust levels \tabularnewline
150   & 0.81 & 14.2 & Normal dust levels \tabularnewline
200   & 1.08 & 11.8 & Normal dust levels \tabularnewline
300   & 1.62 & - & Normal dust levels, did not collapse
\end{tabular}
\caption{Results from \unit{88}{\Msol} collapses of uniform gas}
\label{tbl:Results88MsolCollapse}
\end{table}

\begin{table}
\centering
\begin{tabular}{c|c|c|>{\raggedright}p{3cm}}
$T_{\textrm{gas}} / \kelvin$ & $\alpha$ &  $M_* / \Msol$   &  Comments \tabularnewline
\hline
 10   &  0.003 & 4809  &  No thermalised field \tabularnewline
200   &  0.067 & 3976  &  No thermalised field \tabularnewline
 10   & 0.003 & 4368  & \tabularnewline
200   & 0.067 & 2373  & \tabularnewline
\end{tabular}
\caption{Results from \unit{5630}{\Msol} collapses of uniform gas}
\label{tbl:Results5630MsolCollapse}
\end{table}

\begin{figure}
\centering
\includegraphics[scale=0.4]{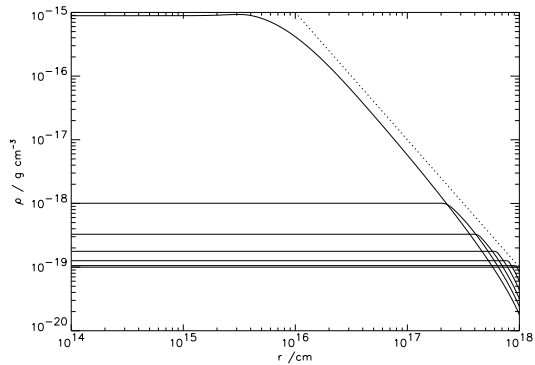}
\caption{Initial collapse of \unit{200}{\kelvin} \unit{241}{\Msol} cloud.
The curves are plotted for $t=0$ and after 0.17, 0.34, 0.51, 0.69, 0.84, and 1.0 freefall
times.
Dotted line is $r^{-2}$ curve}
\label{fig:EarlyWarmUniformCollapse}
\end{figure}

\begin{figure}
\centering
\includegraphics[scale=0.4]{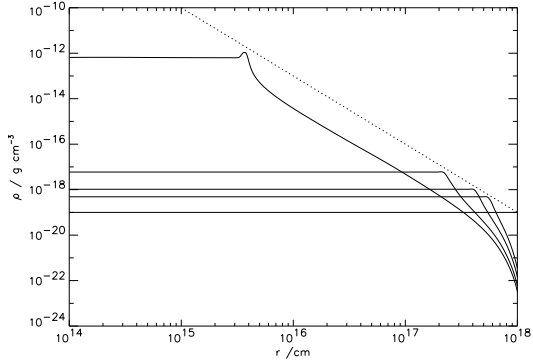}
\caption{Initial collapse of \unit{10}{\kelvin} \unit{241}{\Msol} cloud.
The curves are plotted for $t=0$ and after 0.76, 0.84, 0.94 and 1.0 freefall times.
Dotted line is $r^{-3}$ curve}
\label{fig:EarlyColdUniformCollapse}
\end{figure}

\begin{figure}
\centering
\includegraphics[scale=0.4]{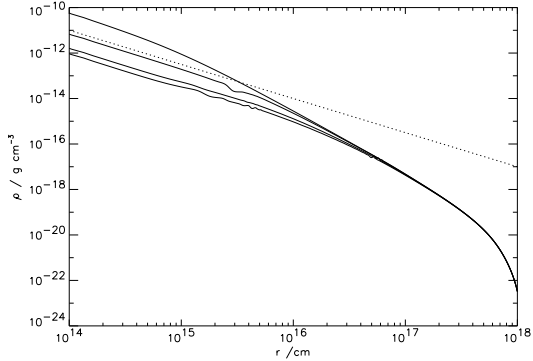}
\caption{Middle phase of collapse of \unit{10}{\kelvin} \unit{241}{\Msol} cloud.
The core grew from \unit{101}{\Msol} to \unit{156}{\Msol} during the period
of \unit{381}{\yyear} plotted.
Dotted line is $r^{-1.5}$ curve}
\label{fig:MidColdUniformCollapse}
\end{figure}

Figures~\ref{fig:EarlyWarmUniformCollapse}, \ref{fig:EarlyColdUniformCollapse},
\ref{fig:MidColdUniformCollapse} and~\ref{fig:TerminalColdUniformCollapse}
depict snapshots of the flow profile and illustrate the chief phases of the
collapse of homogeneous clouds subject to radiative feedback.
In all cases, a core forms after a free-fall time, where
\begin{equation*}
t_{\textrm{ff}} = \sqrt{ \frac{3 \pi}{32 G \rho_{0}} }
= \unit{\scinot{2.1}{5} \parenfrac{\rho_{0}}{\unit{10^{-19}}{\grampercubiccmnp}}^{-\frac{1}{2}}}{\yyear}
\end{equation*}
which is the standard result for the pressure free collapse of a homogeneous
cloud.
There is obviously no core luminosity (or feedback applied) up to the point
of core formation.
The density profile during the initial collapse depends on the temperature
of the flow:
warmer flows exhibit the behaviour discussed by \citet{1969MNRAS.145..271L}
and \citet{1969MNRAS.144..425P} - a flat central density profile with
$r^{-2}$ wings (Figure~\ref{fig:EarlyWarmUniformCollapse}).
The colder collapses produce larger constant density regions as they approach
homologous collapse, with steeper wings (typically closer to $r^{-3}$, see
Figure~\ref{fig:EarlyColdUniformCollapse}).

\begin{figure}
\centering
\includegraphics[scale=0.4]{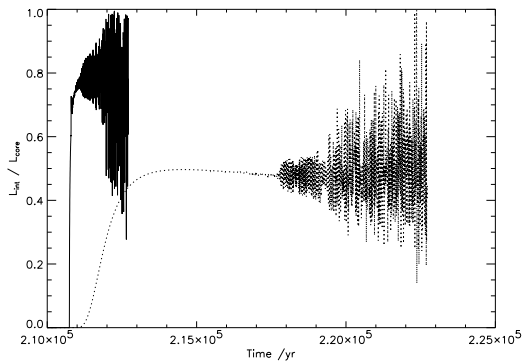}
\caption{Comparison of intrinsic (nuclear) luminosity to total core luminosity for
\unit{10}{\kelvin} (solid line) and \unit{200}{\kelvin} (dotted line) \unit{241}{\Msol} clouds}
\label{fig:LintLcoreUniformCompare}
\end{figure}

\begin{figure}
\centering
\includegraphics[scale=0.4]{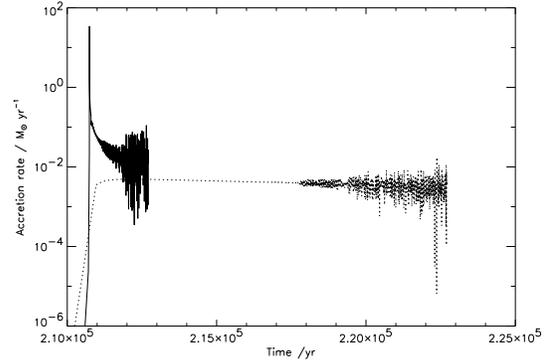}
\caption{Accretion rates for \unit{10}{\kelvin} and \unit{200}{\kelvin} \unit{241}{\Msol} clouds.
Solid line is for \unit{10}{\kelvin} gas,
dotted line follows \unit{200}{\kelvin} cloud}
\label{fig:AccretionRatesUniform}
\end{figure}

\begin{figure}
\centering
\includegraphics[scale=0.4]{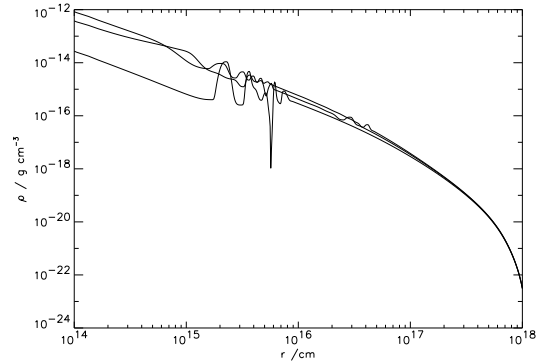}
\caption{Terminal phase of collapse of \unit{10}{\kelvin} \unit{241}{\Msol} cloud.
This plot covers a period of \unit{1200}{\yyear} at the end of the simulation}
\label{fig:TerminalColdUniformCollapse}
\end{figure}

Once the core has formed, the inner region of the flow, which is dominated
by the core's gravity, approaches the $r^{-1.5}$ profile expected for steady
flow onto a point mass (Figure~\ref{fig:MidColdUniformCollapse}).
At this stage, the dominant contribution to the core luminosity is initially
from accretion, but the relative importance of the intrinsic core luminosity
increases as the core mass grows (Figure~\ref{fig:LintLcoreUniformCompare}).
In the final phase of the evolution, the accretion flow starts to be modulated
by the effect of radiative feedback on the dust - oscillations in the
accretion rate may be seen in Figure~\ref{fig:AccretionRatesUniform}.
Radiation pressure temporarily reduces the flow rate from the dust destruction
radius, which produces a drop in the accretion rate onto the core after one
freefall time from $r_{\textrm{d}}$.
The corresponding drop in luminosity causes a reduction in feedback at
$r_{\textrm{d}}$, and the accretion rate climbs again.
The effect of this time delayed feedback is imprinted as oscillations in the
density profile (Figure~\ref{fig:TerminalColdUniformCollapse}).
In the last profile, the large drop in density at the inner edge of the dust
shell is plainly visible, as the flow is effectively stalled by the
momentum imparted by the photon field.

At a quantitative level, the behaviour of the flow can be understood in
terms of the analytic expressions presented in Appendix~\ref{sec:scaling}.
In particular, accretion is found to stall  at the point where the momentum
flux in the accretion flow at $r_{\textrm{d}}$ matches $L/c$ from the core.
The evolution of the dust destruction radius (computed in the code according
to the algorithm of section~\ref{sec:SimpleRadTrans}) may be compared with
the approximation $L = A r_{\textrm{d}}^2$ made in the Appendix, and was
found to be in satisfactory agreement over most of the evolution.
Values of $A$ were found to be similar to those of \citet{1987ApJ...319..850W}.

\begin{figure}
\centering
\includegraphics[scale=0.4]{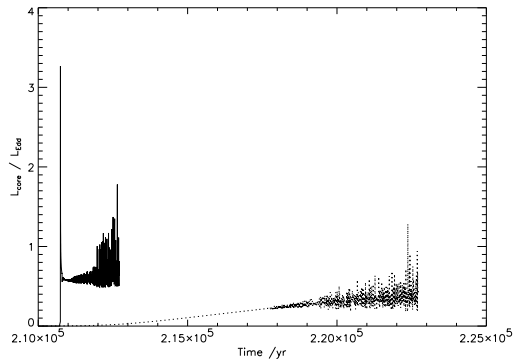}
\caption{Comparison of total luminosity to Eddington luminosity for
\unit{10}{\kelvin} (solid line) and \unit{200}{\kelvin} (dotted line) \unit{241}{\Msol} clouds}
\label{fig:LcoreLeddUniformCompare}
\end{figure}

As Figure~\ref{fig:AccretionRatesUniform} demonstrates, the colder gas clouds
underwent a spike of accretion as shells from a range of radii arrive almost
simultaneously at the centre (as expected for a homogeneous collapse).
Roughly half the mass of the cloud goes into the core during this first spike.
The associated accretion rate is high, and briefly super-Eddington - see
Figure~\ref{fig:LcoreLeddUniformCompare}.\footnote{Note that our simulation
takes no account of electron scattering opacity, since the ionised regions
of the flow are not resolved.
However, since the integrated energy input during the super-Eddington phase
is rather less than the binding energy of the core, the core would not be
liable to disruption}
For a star of mass \unit{M}{\Msol}, the Eddington Luminosity is
\unit{\scinot{3}{4}M}{\Lsol}.
After the initial accretion spike, accretion continued but soon large
oscillations develop in the accretion rate, and accretion subsequently stalls.

The `warm' run behaves somewhat differently.
The accretion rate rises more slowly, and remains steady for about $10\%$ of
the freefall time of the original cloud.
The rising intrinsic luminosity of the core causes feedback to play an
increasing role.
Oscillations develop in the accretion rate, leading to the eventual stalling
of the accretion flow.


\subsection{Collapse from Power Law Initial Conditions}

We now consider collapses in which the initial density profile scales
as $r^{-1}$.
Final stellar masses as a function of $\alpha$ (still relevant for
this density distribution) are given in
Table~\ref{tbl:Results262MsolRhoInvRadCollapse}.
The densities were normalised so that models of given mass and temperature
had $\alpha$ values similar to those starting from uniform conditions.

\begin{table}
\centering
\begin{tabular}{c|c|c|>{\raggedright}p{3cm}}
$T_{\textrm{gas}} / \kelvin$  &  $\alpha$  &  $M_* / \Msol$   &  Comments \tabularnewline
\hline
10   &  0.02  & 25.3 & \tabularnewline
20   &  0.05  & 28.0 & \tabularnewline
50   &  0.11  & 28.2 & \tabularnewline
100  &  0.23  & 36.3 & \tabularnewline
150  &  0.35  & 51.4 & \tabularnewline
200  &  0.46  & 64.8 & \tabularnewline
300  &  0.68  & 53.2 & \tabularnewline
400  &  0.92  & 45.7 & \tabularnewline
10   &  0.02  & 14.3 & Normal dust levels \tabularnewline
20   &  0.05  & 15.8 & Normal dust levels \tabularnewline
50   &  0.11  & 15.4 & Normal dust levels \tabularnewline
100  &  0.23  & 11.5 & Normal dust levels \tabularnewline
150  &  0.35  & 12.0 & Normal dust levels \tabularnewline
200  &  0.46  & 11.3 & Normal dust levels \tabularnewline
300  &  0.68  & 11.5 & Normal dust levels \tabularnewline
400  &  0.92  & 11.7 & Normal dust levels
\end{tabular}
\caption{Results from \unit{262}{\Msol} collapses of $\rho \propto r^{-1}$ gas}
\label{tbl:Results262MsolRhoInvRadCollapse}
\end{table}

\begin{figure}
\centering
\includegraphics[scale=0.4]{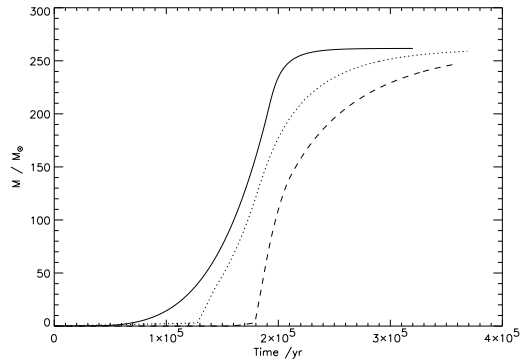}
\caption{Accretion histories for cloud started from $r^{-1}$ initial density profile and no feedback.
Solid curve is for \unit{10}{\kelvin} gas,
the dotted line \unit{100}{\kelvin},
and the dashed line \unit{200}{\kelvin} gas}
\label{fig:AccreteHistoryPowerLawNoLum}
\end{figure}

\begin{figure}
\centering
\includegraphics[scale=0.4]{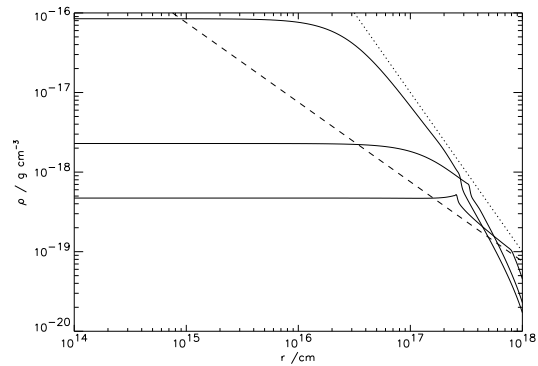}
\caption{Collapse of \unit{200}{\kelvin} cloud started from $r^{-1}$ initial density profile.
The solid curves are plotted for 0.36, 0.72, and 0.86 freefall times from the
edge of the simulation.
The dashed line is the initial density profile.
Dotted line is $r^{-2}$ curve}
\label{fig:EarlyWarmPowerLawCollapse}
\end{figure}

The effect of feedback may be most readily understood by first considering
the growth of the central core in such models when feedback is \emph{not}
included.
A set of sample curves is presented in
Figure~\ref{fig:AccreteHistoryPowerLawNoLum}.
The core growth of the coldest run roughly follows the $M \propto t^4$ law
derived by considering the freefall collapse of thin shells.
The core is therefore built over a much longer interval than the homogeneous
case.
The warmer runs behave in a manner similar to their homogeneous counterparts,
since early core formation is suppressed by the action of pressure gradients.
Since $M(r) \propto r^2$, the gravitational acceleration, $g$, is constant
throughout the initial cloud.
However, the acceleration due to the pressure gradient varies according to
\begin{equation*}
\frac{1}{\rho} \cdot \deriv{P}{r} \sim
\frac{1}{\rho} \cdot \frac{P}{r} \sim \frac{c_{\textrm{iso}}^2}{r} \sim r^{-1}
\end{equation*}
so pressure support improves towards the centre of the cloud.
Consequently, the gas evolved towards the Larson--Penston solution initially,
as maybe seen from Figure~\ref{fig:EarlyWarmPowerLawCollapse}.
The subsequent behaviour of the warmer clouds closely mirrored those collapsing
from uniform initial conditions.

When feedback was included, it was found that the warmest runs led to final
stellar masses almost indistinguishable from the corresponding homogeneous
simulations (c.f. Tables~\ref{tbl:Results241MsolCollapse}
and~\ref{tbl:Results262MsolRhoInvRadCollapse}).
This is as one would expect, given the appearance of the Larson--Penston solution
prior to core formation.
However, for still lower temperatures ($T_{\textrm{gas}} < \unit{200}{\kelvin}$, or
$\alpha < 0.5$ for the conditions considered here), the trend is reversed, and
final stellar mass \emph{decreases} with lower temperatures.
This behaviour is explained by Figure~\ref{fig:AccreteHistoryPowerLawNoLum}.
The lack of pressure support caused the cores to form over a \emph{longer} period
of time, since the formation of the core is not delayed by pressure gradients.
This permits feedback to operate when a substantial fraction of the initial cloud
mass still lies beyond $r_{\textrm{d}}$.

The interaction of the collapse with the radiative feedback followed
a similar pattern to that observed in the collapses from uniform
clouds.
The collapses with feedback closely followed the behaviour of those
without, until the central luminosity became significant.
Then, the flow began to oscillate on a short timescale, and accretion
halted rapidly.


\section{Discussion}
\label{sec:discussion}

Three phases of collapse were observed, where a different physical
effect was dominant:
\begin{enumerate}
\item Self-gravity of the cloud
      \label{stage:SelfGrav}
\item Gravity of the core
      \label{stage:CoreGrav}
\item Luminosity of the core
      \label{stage:CoreLum}
\end{enumerate}
Stages~\ref{stage:SelfGrav} and~\ref{stage:CoreGrav} have been
studied for a number of years.
Radiative feedback is important in the final phase.

A cursory glance at Tables~\ref{tbl:Results241MsolCollapse},
\ref{tbl:Results88MsolCollapse} and~\ref{tbl:Results5630MsolCollapse}
suggests that, given the right initial conditions, massive stars may
form despite the effects of radiative feedback.
For example, Table~\ref{tbl:Results5630MsolCollapse} documents the
formation of stars containing thousands of solar masses of material;
we cannot of course rule out the possibility that such objects would
fragment into a cluster of lower mass stars, but this is not a
radiative effect.
Such stars are not observed, but we find no reason why radiative
feedback should prevent their formation.
Higher initial cloud masses lead to higher final stellar masses.
However, we do find that values of $\alpha$
(defined in equation~\ref{eq:DefineAlpha}) close to unity cause the
fraction of the cloud mass accreted to be very low, whereas small
$\alpha$ values gave rise to a high accretion fraction.
This variation with $\alpha$ arises because feedback can only become effective
in stage~\ref{stage:CoreLum} of the collapse.
For the colder homogeneous clouds, most of the mass had already arrived within
$r_{\textrm{d}}$ by the time this stage was reached.
By contrast, warmer collapses possessed pressure gradients which modified
the density profile, and caused the core to form over a more extended
period.
Indeed, the steady accretion rate of the warm cloud depicted in
Figure~\ref{fig:AccretionRatesUniform} is rather similar to the
conditions considered by \citet{1987ApJ...319..850W}, although
our differing parameterisations of the protostellar core prevent
quantitative comparison.
By contrast, the cold collapse is \emph{qualitatively} different
and points to the possibility that massive star formation may not
be as difficult as previously envisaged.

\citet{1987ApJ...319..850W} concluded that depleted dust abundances
were necessary to form massive stars, yet the current work has demonstrated
the formation of stars $> \unit{100}{\Msol}$ for the
\citet{1984ApJ...285...89D} dust abundances.
However, the condition derived by \citeauthor{1987ApJ...319..850W} to support
this requirement was analogous to the Eddington Limit, and hence was based
on accelerations.
Balancing accelerations is only useful when the processes involved
are close to equilibrium (e.g. radial motion through an accretion disc).
The present work studied highly dynamical collapses, and the velocities
obtained in the early stages of the collapse were easily sufficient to
overcome the net outward acceleration as the luminosity rose.

\begin{figure*}
\centering
\includegraphics[scale=0.8]{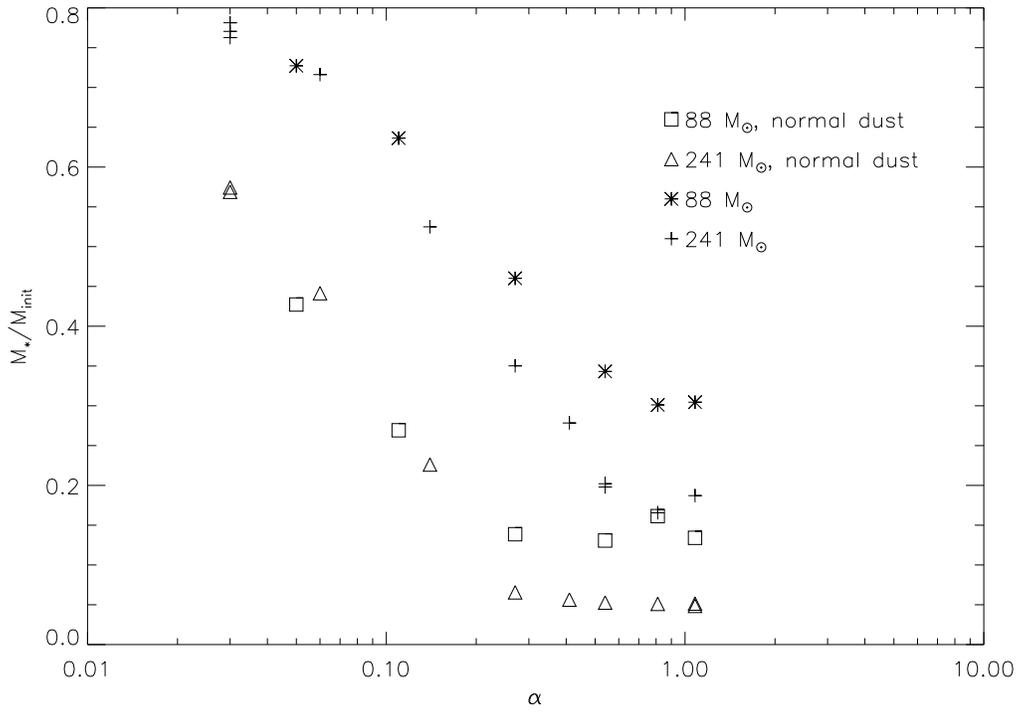}
\caption{Initial $\alpha$ against final fraction of mass accreted for collapses from a uniform cloud.
Note that $N_{\textrm{J}} \approx \alpha^{-\frac{3}{2}}$ for the uniform collapses}
\label{fig:InitNJagainstFinalAccreteFracUniformCollapse}
\end{figure*}

A plot of initial $\alpha$ against the fraction of mass accreted onto
the star (a measure of the efficiency) for an initially uniform cloud is
presented in Figure~\ref{fig:InitNJagainstFinalAccreteFracUniformCollapse}.
In general, lower $\alpha$ values lead to more massive stars.
For $\alpha \sim 1$ and higher dust opacities, the final stellar mass
obtained seems to tend to a constant value
(c.f. Tables~\ref{tbl:Results241MsolCollapse}
and~\ref{tbl:Results88MsolCollapse}), possibly indicating that a true
radiation pressure limit is being found.
For the depleted dust mixture, the stars obtained can still be quite
massive -- $>\unit{40}{\Msol}$.

\begin{figure*}
\centering
\includegraphics[scale=0.8]{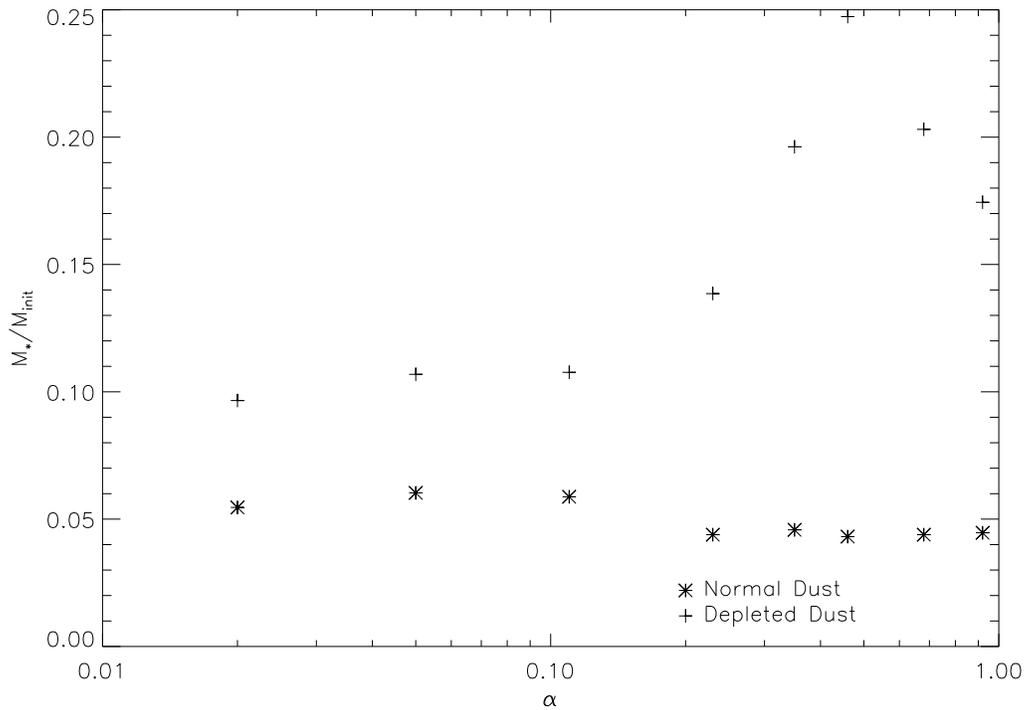}
\caption{Initial $\alpha$ against final fraction of mass accreted for
collapses from an initial density profile of $\rho \propto r^{-1}$.
In all cases, the total cloud mass was \unit{262}{\Msol}
}
\label{fig:InitAlphaagainstFinalAccreteFracPowerLawCollapse}
\end{figure*}

The results from collapses of $\rho \propto r^{-1}$ gas, as presented in
Table~\ref{tbl:Results262MsolRhoInvRadCollapse} show that radiative feedback
can be very important under more condensed initial conditions.
A plot of $\alpha$ against the star formation efficiency is presented in
Figure~\ref{fig:InitAlphaagainstFinalAccreteFracPowerLawCollapse}.
The rise in efficiency with increasing temperature is plain for the depleted
dust, as is the steady value obtained for normal dust abundances.
In the case of normal dust abundances, a limit of $\sim \unit{12}{\Msol}$ is
implied, with the variations around this value not being very significant.
For the depleted dust mixture, the rather surprising result that colder gas does
not lead to more massive stars is related to the behaviour of the
collapse before radiative feedback becomes significant.
If thermal effects are significant, they can delay core formation, and hence the
onset of radiative feedback.
During this delay, gas settled inside the dust destruction radius (see
Figure~\ref{fig:AccreteHistoryPowerLawNoLum} and associated discussion).

\section{Conclusion}
\label{sec:conclude}

We have developed a new simplified algorithm for calculating radiative feedback
in hydrodynamic simulations.
This method combines the computational economy of a frequency averaged
treatment of the thermalised field (via a modified diffusion approximation) with
the benefits of a frequency dependent treatment of the attenuated stellar
field.
This latter feature ensures that the flow receives the sharp impulse at the
dust destruction radius that is found in full radiative transfer calculations;
we have demonstrated (Section~\ref{sec:SimpleRadTrans}) that this algorithm is
significantly superior to methods using grey opacities, which have become
the standard method in hydrodynamical simulations to date.

We have applied this algorithm to the formation of massive stars by spherically
symmetric accretion.
Although this obviously does not represent a realistic geometry for OB star
formation, our motivation has been to compare the hydrodynamical treatment of
this process with the steady state models of \citet{1987ApJ...319..850W}, in
order to evaluate how the time-dependence of the hydrodynamic calculation
affects the results.
Since we find (see below) that massive star formation is rather unproblematical
(given appropriate initial conditions), we would expect it to be yet easier
with a realistic (disc) accretion geometry.
Our chief conclusions are as follows:
\begin{enumerate}
\item Massive star formation is favoured by models in which the luminous core
forms \emph{quickly}.
In this case, much of the cloud is already within the dust destruction radius
(and hence immune from feedback) when the core luminosity becomes significant.
The most favourable conditions involve cold homogeneous collapse,
since in this case the core assembly time is shortest.
Warmer collapses (with thermal to gravitational energy ratios near unity) produce
lower final stellar masses, but can still readily produce OB stars given
sufficiently massive progenitor clouds.
For these warm collapses, the final stellar masses are rather insensitive to the
initial cloud density profile

\item We find that, in contrast to \citet{1987ApJ...319..850W}, a depleted
dust mixture is \emph{not} necessary for the formation of OB stars.
The reason for this difference is that \citeauthor{1987ApJ...319..850W} required
that the net acceleration at the cloud's outward edge was inward, in order
to attract the initially stationary material.
In the hydrodynamic collapses, the fluid is already inflowing at close to its
freefall velocity by the time the core luminosity becomes significant.
However, increasing the optical depth implies a stronger deceleration of the
flow from the diffuse field, which does reduce the final stellar masses.
\end{enumerate}

Although, as stated above, we do not consider our simulations to be a realistic
portrait of massive star formation, we end with a few remarks about the
possible relevance of these findings to OB star formation.
Firstly, we might expect that the qualitative conclusion that massive star
formation is favoured by rapid core formation to hold in other geometries.
Such rapid formation is promoted by rather homogeneous, highly Jeans unstable
conditions -- as might be expected to arise in regions of triggered star
formation (see, e.g. \citet{1997A&A...328..167B} or
\citet*{2001AJ....121.3075O}).
Secondly, we note that in our models in which feedback effects were
particularly severe (either due to warm initial conditions or the use of a
standard grain opacity), the resulting stellar mass often ended up near
to $\sim \unit{12}{\Msol}$.
The lack of any obvious signature in the IMF around this value therefore
argues that the dominant population of progenitor clouds does not fall into
this category.


\appendix

\section{Scaling Relations}
\label{sec:scaling}

For accretion to occur, despite radiation pressure on the
embedded dust grains, the ram pressure at the dust
destruction radius, $r_{\textrm{d}}$ must exceed the momentum of the
radiation field.
That is
\begin{equation}
\dot{M} u_{\textrm{d}} > \frac{L}{c}
\label{eq:RamPressureLumBalance}
\end{equation}
where $u_{\textrm{d}}$ is the inflow velocity at $r_{\textrm{d}}$.
This is the condition used by \citet{1987ApJ...319..850W} at the
inner edge of their dust shell.

The value of $u_{\textrm{d}}$ may be calculated on the assumption that the stellar
mass dominates, and that the material is in free fall at the destruction
radius.
To make further progress, assumptions must be made about the source of
the luminosity.


\subsection{Luminosity Dominated by Accretion}

If the core's luminosity is dominated by accretion, one obtains
\begin{equation*}
\dot{M} \sqrt{2 G M_*} > \frac{G \dot{M} M_*}{c R_*} r_{\textrm{d}}^{\frac{1}{2}}
\end{equation*}

To compute the accretion luminosity, a mass-radius relation is required.
A simple mass radius relation for massive stars is
\begin{equation*}
R_* = \parenfrac{M_*}{\Msol}^{\frac{3}{5}} \Rsol
\end{equation*}
which leads to
\begin{equation*}
r_{\textrm{d}} < \frac{2 c^2 \Rsol^2}{G \Msol^{\frac{6}{5}}} M_*^{\frac{1}{5}}
\end{equation*}
as the condition for accretion.
That is, the dust destruction radius must be \emph{less} than some
\emph{increasing} function of the stellar mass.
This would mean that accretion becomes \emph{easier} over time, and
that radiative feedback would place no limit on stellar masses.
Evaluating the constants gives the following prediction for steady state
accretion to occur:
\begin{equation}
r_{\textrm{d}} < \unit{4400 \parenfrac{M_*}{\Msol}^{\frac{1}{5}}}{\AU}
\label{eq:DustRadiusPrediction}
\end{equation}

Of course, the dust destruction radius, $r_d$ varies with core
luminosity.
As a first approximation, the dust destruction radius will vary
as
\begin{equation}
L = A r_{\textrm{d}}^2
\label{eq:DustDestructRadApprox}
\end{equation}
for some constant $A$.
Table~3 of \citet{1987ApJ...319..850W} suggests that
$A \approx \unit{\scinot{3}{8}}{\erg\usk\reciprocal\second\usk\persqcmnp}$.
If it is assumed that equation~\ref{eq:DustDestructRadApprox}
is an expression of the Stefan--Boltzmann Law, then the implied
temperature is less than \unit{1000}{\kelvin} - far below the
nominal dust melting temperature, and another example of the importance
of the wavelength dependence of the dust opacity.
However, with this dependence folded into $A$,
equation~\ref{eq:DustDestructRadApprox} may be safely used to estimate
the value of $r_{\textrm{d}}$.

Putting this together with the free fall assumption yields
\begin{equation*}
\dot{M} \sqrt{2 G M_*} > \frac{L^{\frac{5}{4}}}{c \sqrt[4]{A}}
\end{equation*}
or
\begin{equation}
4 \dot{M}^4 G^2 M_*^2 > \frac{L^5}{A c^4}
\label{eq:BasicAccCondition}
\end{equation}
Substituting an accretion luminosity then yields
\begin{equation*}
\dot{M} < 4 G^{-3} M_*^{-3} R_*^5 A c^4
\end{equation*}
Finally, putting in the mass-radius relation gives
\begin{equation}
\dot{M} < \frac{4 A \Rsol^5 c^4}{G^3 \Msol^{3}}
          \sim \unit{6}{\Msol\usk\reciprocal\yyear}
\label{eq:AccLimitAccLum}
\end{equation}
as the condition for accretion.
Rather surprisingly, this is independent of the stellar mass.
However, no star of reasonable mass could accrete at this
rate and remain on the main sequence.
We therefore conclude that accretion luminosity on its own
is unlikely to shut down an accretion flow.


\subsection{Luminosity Dominated by Nuclear Burning}

If the luminosity is dominated by nuclear reactions in the
core, a different relation between $L$ and $M$ is obtained.
For main sequence stars, $L \sim M^{\psi}$, where
$\psi$ is in the range of two to three for most stars
(although the most massive stars have $\psi \sim 1$).
Substituting the power law relation into equation~\ref{eq:BasicAccCondition}
suggests that the condition on the accretion rate will become
\begin{equation}
\dot{M}^4 > \frac{\Lsol^5}{4 G^2 A c^4 \Msol^{5 \psi}} M_*^{5 \psi - 2}
\label{eq:AccLimitIntrinsicLum}
\end{equation}
Within the likely range of values for $\psi$, this \emph{minimum} accretion
rate is a rather rapidly increasing function of $M_*$.
Hence, accretion is becoming harder, and is likely to shut down.


\section*{Acknowledgments}

The authors would like to thank Jim Pringle and Ian Bonnell for several useful discussions.
RGE is also particularly grateful to Matthew Bate, for his aid with {\sc ZEUS}.
Chris Tout kindly supplied his ZAMS formulae and information on possible accretion rates.


\bibliography{bibs/radiativetrans,bibs/dust,bibs/zeus,bibs/observations,bibs/hydro,bibs/compute,bibs/stellarevolve,bibs/accretiondisks,bibs/starform}
\bibliographystyle{mn2e}

\bsp

\label{lastpage}

\end{document}